\documentclass[twocolumn,showpacs,showkeys,superscriptaddress]{revtex4}
\usepackage{graphicx}
\usepackage{bm}
\usepackage{color}
\usepackage{amsmath}
\usepackage{natbib}

\begin{document}


\title{The effect of spin magnetization in the damping of electron plasma oscillations}

\author{Pablo S. Moya}
\email{pmoya@levlan.ciencias.uchile.cl}
\affiliation{Departamento de F\'\i sica, Facultad de Ciencias, Universidad de Chile, Casilla 653, Santiago, Chile.}

\author{Felipe A. Asenjo}
\email{fasenjo@levlan.ciencias.uchile.cl}
\affiliation{Departamento de F\'\i sica, Facultad de Ciencias, Universidad de Chile, Casilla 653, Santiago, Chile.}
\affiliation{Departamento de Ciencias, Facultad de Artes Liberales,  Universidad Adolfo Ib\'a\~nez, Diagonal Las Torres 2640, Pe\~nalol\'en, Santiago, Chile.}

\date{\today}

\begin{abstract}
The effect of spin of particles in the propagation of plasma waves is
studied using a semi-classical kinetic theory for a magnetized
plasma. We focus in the simple damping effects for the electrostatic
wave modes besides Landau damping. Without taking into account more
quantum effects than spin contribution to Vlasov's equation, we show
that spin produces a new damping or instability which is proportional
to the zeroth order magnetization of the system. This correction
depends on the electromagnetic part of the wave which is coupled with
the spin vector.
\end{abstract}

\pacs{03.65.Sq, 52.25.Dg, 52.25.Xz, 52.27.Gr, 52.35.Fp}
\keywords{Spin plasma dynamics, Landau damping, spin magnetization}

\maketitle


One of the most important physical results in the propagation of
plasma waves which cannot be deduced by a fluid description is Landau
damping \cite{landau}. This effect predicts that an electron plasma
wave, in an collisionless plasma,
suffers damping owing to the wave-particle interaction. Damping
mechanism will depend on the velocity distribution of the particles
which are often Maxwellian. 
The mathematical procedures to obtain Landau damping are
very standard in kinetic theory. Thus, the formalism has been extended
for other kind of electron interactions, for example thermal ion
Landau damping effects~\cite{goldston}, and Landau damped electron
waves by photons \cite{bing} or neutrinos \cite{losilva}.

On the other hand, recently there have been a huge interest in the field of
plasma physics for the dynamics of the spin of electrons, and how its
quantum nature affects the different known modes of propagations
\cite{shukla, misra,shukla2} or other
properties~\cite{gosh,eliason}. These treatments are useful in, for
example, astrophysical systems \cite{baring} or high-energy lasers
\cite{kremp}. Often these effects are important in high density,
low temperature or strong magnetic fields conditions, but it has
been shown that for some systems at high temperaure the spin dynamics
play a crucial role \cite{manfredi}.

In a previous work~\cite{faz}, we present a first approach to
calculate a correction to Landau damping due to spin. In this letter,
we complete the analysis done in the previous work finding a full
solution for the damping produced by the spin to electrostatic
modes. We will show that this new damping is proportional to the spin
magnetization of the plasma, and it depends on the electromagnetic
part of the wave.

To obtain the correction to the Landau damping produced by the spin of
the plasma constituents, we use the semi-classical kinetic theory
constructed in Ref.~\cite{brodin} which start from the Pauli
Hamiltonian. Here, the dynamics of the spin of particles is included
in a Vlasov equation for a generalized distribution function. Other
quantum corrections, as Bohm potential, spin-spin interaction force
and high order force terms in the spin evolution equation are
neglected. Thus, Vlasov equation, for 
particles with velocity ${\bf v}$, spin vector ${\bf s}$ and with
distribution function $f=f({\bf r}, {\bf v}, {\bf s}, t)$ is
\begin{eqnarray}
\frac{\partial f}{\partial t}&+&{\bf v}\cdot\nabla f+\left[\frac{q}{m}\left({\bf E}+{\bf
      v}\times{\bf B}\right)+\frac{2\mu_e}{m\hbar}\nabla\left({\bf s}\cdot{\bf
      B}\right)\right]\cdot\nabla_v f\nonumber\\
&&\qquad\qquad\qquad\qquad+\frac{2\mu_e}{\hbar}\left({\bf s}\times{\bf B}\right)\cdot\nabla_s f=0\, ,
\label{vlasov}
\end{eqnarray}
where $q=-e$ and $m$ are the electron charge and mass respectively, $\mu_e=-g
e\hbar /(4m)$ is the electron magnetic moment and $g\approx 2.002319$ is the
electron spin factor. The Fermi-Dirac equilibrium distribution
function $f_0({\bf v}, {\bf s})$ is
\cite{brodin}
\begin{equation}
 f_0({\bf v}, {\bf s})=\frac{B_0 \mu_e{\tilde f}_0({\bf v})}{4\pi k_B
   T\sinh (B_0 \mu_e/k_B T)}\exp\left(\frac{2\mu_e {\bf s}\cdot{\bf B}_0}{\hbar
     k_B T}\right)\, ,
\label{distribu0}
\end{equation}
where $T$ is the temperature, $k_B$ is the Boltzmann constant, ${\bf
  B}_0$ is a background magnetic field ($B_0=|{\bf B}_0|$) and
${\tilde f}_0$ is the classical Maxwellian distribution function 
\begin{equation}
{\tilde f}_0({\bf v})= \left(\frac{m}{2\pi k_B T}\right)^{3/2}\exp\left(-\frac{m v^2}{2 k_B T}\right)\, ,
\label{distrMaxwe}
\end{equation}
with $v=|{\bf v}|$. The distribution function \eqref{distribu0} is
normalized as $\int f_0 d{\bf v} d{\bf s}=1$, where
the integration is made over the three degree of freedom in velocity
space and the two degree of freedom in spin space.

Now, we use the kinetic formalism \eqref{vlasov} and a similar
analysis of Ref.~\cite{faz} to derive the dispersion
relation for electron plasma oscillations in a magnetized
plasma which interacts with the spin of the particles. The electric and magnetic
fields will be perturbed in the form ${\bf E}={\bf E}_1$ and ${\bf B}={\bf B}_0+{\bf B}_1$
respectively. The terms with subscript $0$ are the zeroth order
equilibrium quantities, and the terms with subscript $1$ are the first
order perturbed quantitites. The distribution function is perturbed as $f({\bf r}, {\bf v},
{\bf s}, t)=f_0({\bf v}, {\bf s})+{\hat f}_1({\bf r}, {\bf v}, {\bf
  s}, t)$ and we choose ${\bf B}_0=B_0\hat z$. The perturbed distribution ${\hat f}_1$ will have the form ${\hat f}_1({\bf
  r}, {\bf v}, {\bf s}, t)=f_1({\bf v}, {\bf s})\exp(i{\bf k}\cdot{\bf
  r}-i\omega t)$, with similar assumption for other perturbed
quantities. Here, ${\bf k}$ and $\omega$ are the wavenumber and the
frequency of the wave.

Linearizing Eq.~\eqref{vlasov}, with the velocity ${\bf v}$ and spin ${\bf s}$
as independet variables and following Ref.~\cite{faz}, we can find the perturbed distribution
function as
\begin{eqnarray}
 f_1&=&\frac{-i}{\omega-{\bf k}\cdot{\bf v}}\left(\frac{q}{m}{\bf E}_1\cdot\nabla_v f_0\right.\nonumber\\
&&\left.+\frac{2\mu_e}{m\hbar}\nabla\left({\bf s}\cdot{\bf B}_1\right)\cdot\nabla_v
  f_0+\frac{2\mu_e}{\hbar}\left({\bf s}\times{\bf B}_1\right)\cdot\nabla_s f_0\right)\,
,\nonumber\\
&&
\label{f1}
\end{eqnarray}
because ${\bf v}\times{\bf B}_1\cdot\nabla_v f_0=0$.

From here, we are going to focus in the study of spin correction to
the Landau damping for electrostatic modes. Let us concentrate the
charge density $\rho$ of the plasma which is given by $\rho=qn_0\int f_1 d{\bf
  v}d{\bf s}$, where $n_0$ is the equilibrium density. Using the
Maxwell equation ${\bf k}\times{\bf E}_1=\omega{\bf B}_1$, the perturbed
distribution function \eqref{f1},  and defining the quantities ${\bf
  E}_\perp={\bf k}\times{\bf E}_1/k$ and $E_\parallel={\bf k}\cdot{\bf E}_1/k$ with
$k=|{\bf k}|$, the charge density becomes in
\begin{eqnarray}
\rho&=&-iqn_0\int d{\bf s}\int_{-\infty}^\infty \frac{d{\bf v}}{\omega-{\bf k}\cdot{\bf v}}\left(\frac{q}{mk}E_\parallel\left({\bf k}\cdot\nabla_v f_0\right)\right.\nonumber\\
&&\left.+\frac{i2\mu_e k}{m\hbar\omega}\left({\bf s}\cdot{\bf E}_\perp\right){\bf k}\cdot\nabla_v
  f_0+\frac{2\mu_ek}{\hbar\omega}\left({\bf s}\times{\bf E}_\perp\right)\cdot\nabla_s f_0\right)\,
.\nonumber\\
&&
\label{f12}
\end{eqnarray}

This charge density must be used in the Poisson's equation $i k E_\parallel=4\pi
\rho$ to obtain the dispersion relation for electrostatic modes. In this
way, the dispersion relation is
\begin{widetext}
\begin{equation}
1=\frac{\omega_p^2}{k^2}\int d{\bf s}\int_{-\infty}^\infty \frac{d{\bf v}}{{\bf k}\cdot{\bf
    v}-\omega}\left(1+\frac{2i\mu_e k^2}{q\hbar\omega}\left(\frac{{\bf s}\cdot{\bf
        E}_\perp}{E_\parallel}\right)\right){\bf k}\cdot\nabla_v f_0
+\frac{\omega_p^2}{k^2}\int d{\bf s}\int_{-\infty}^\infty \frac{d{\bf v}}{{\bf k}\cdot{\bf
    v}-\omega}\left(\frac{2\mu_e k^2 m}{q\hbar\omega}\right)\left(\frac{{\bf s}\times{\bf
      E}_\perp}{E_\parallel}\right)\cdot\nabla_s f_0\, ,
\label{disperrelation}
\end{equation}
\end{widetext}
where $\omega_p^2=4\pi e^2 n_0/m$ is the square of the plasma frequency. When
the spin contribution is neglected ($\mu_e=0$), we reobtain the
classical dispersion relation for electrostatic modes.

To solve the dispersion relation \eqref{disperrelation}, we need to
evaluate the two integral involving the spin contribution. The
integration in the two degree of freedom in spin space is done in
spherical coordinates such that $d{\bf s}\equiv d\Omega_s=d(\cos\theta_s)d\phi_s$ where
the subindex $s$ is for spin coordinates. In the same sense, the spin
vector will be ${\bf s}=-\hbar/2\hat s=-\hbar/2\left(\sin\theta_s\cos\phi_s\hat
  x+\sin\theta_s\sin\phi_s\hat y+\cos\theta_s\hat z\right)$, and $\nabla_s f_0=2\mu_e
B_0\sin\theta_s f_0/(\hbar k_B T)\hat \theta$. The choice on the spin orientation
is to minimize the magnetic moment energy, which is consistent with
paramagnetism~\cite{brodin2}. On the other hand, the above
integrations in velocity and spin space can be simplified introducing
the one-dimensional distribution
\begin{equation}	
 F_0(u,{\bf s})\equiv \int f_0 \ \delta\left(u-\frac{{\bf k}\cdot{\bf v}}{k}\right)d{\bf v}\, .
\label{F0}
\end{equation}

We can use \eqref{F0} to rewrite the dispersion relation \eqref{disperrelation} as
\begin{widetext}
\begin{equation}
1=\frac{\omega_p^2}{k^2}\int d{\Omega_s}\int_{-\infty}^\infty \frac{d
  u}{u-\omega/k}\left(1-\frac{i\mu_e k^2}{q\omega}\left(\frac{{\hat s}\cdot{\bf
        E}_\perp}{E_\parallel}\right)\right)\frac{\partial F_0}{\partial u}
-\frac{2\omega_p^2\mu_e^2 m B_0}{q\hbar\omega k_B T}\int d{\Omega_s}\int_{-\infty}^\infty \frac{d u \sin\theta_s F_0}{u-\omega/k}\frac{\left({\hat s}\times{\bf E}_\perp\right)\cdot\hat\theta}{E_\parallel}\, .
\label{disperrelation2}
\end{equation}
\end{widetext}

The integrals in Eq.~\eqref{disperrelation2} must be evaluated as a
contour integral considering the singularity at $u_\phi\equiv\omega/k$. This is the
origin of classical Landau damping. We consider the
case of large phase velocity $u_\phi$ and weak damping, where the pole
lies near the real $u$ axis. In this case, $F_0$ and $\partial{F}_0/\partial u$ are
both small near $u_\phi$. Neglecting the thermal correction to  the real part of the
frequency, the first two integrals are given by \cite{chen}
\begin{equation}
\int d{\Omega_s} \int_{-\infty}^\infty \frac{du}{u-\omega/k}\frac{\partial F_0}{\partial u}\simeq \frac{k^2}{\omega^2}+i\pi\left.\frac{\partial{\tilde F}_0}{\partial u}\right|_{u=u_\phi}\, ,
\label{integral1}
\end{equation}
\begin{multline}
\int d{\Omega_s} \int_{-\infty}^\infty \frac{du}{u-\omega/k}\frac{{\hat s}\cdot{\bf
    E}_\perp}{E_\parallel}\frac{\partial F_0}{\partial u}\simeq\\
-\chi\eta\left(\alpha\right)\left(\frac{k^2}{\omega^2}+i\pi\left.\frac{\partial{\tilde F}_0}{\partial u}\right|_{u=u_\phi}\right)\, ,
\label{integral2}
\end{multline}
where $\chi=\hat z\cdot{\bf E}_\perp/E_\parallel$, $\alpha=\mu_e B_0/k_BT$ and
$\eta(x)=\coth(x)-1/x$ is the Langevin function. ${\tilde F}_0$ comes
from \eqref{F0}
and it is defined as
\begin{equation}	
 {\tilde F}_0(u)\equiv \int {\tilde f}_0 \ \delta\left(u-\frac{{\bf k}\cdot{\bf v}}{k}\right)d{\bf v}\, .
 \label{F02}
\end{equation}

The third integral in dispersion relation \eqref{disperrelation2}
vanish due the only relevant spin contribution is anti parallel to the
background magnetic field. Then, the dispersion relation \eqref{disperrelation2} becomes
\begin{equation}
 1=\left(\frac{\omega_p^2}{\omega^2}+\frac{i\pi\omega_p^2}{k^2}\left.\frac{\partial{\tilde
         F}_0}{\partial u}\right|_{u_\phi}\right)\left(1+\frac{i\mu_e
     k^2}{q\omega}\chi\eta(\alpha)\right)\,.
\label{disperrelation3}
\end{equation}

We seek a frequency which has a real and an imaginary part given
by $\omega=\omega_r+i\omega_i$ such that $\omega_i \ll \omega_r $. Using this in
Eq.\eqref{disperrelation3}, and solving for the real and imaginary
parts, we can obtain the frequency for the electrostatic modes
\begin{equation}
 \omega=\omega_r\left(1+\frac{i\pi \omega_r^2}{2 k^2} \left.\frac{\partial{\tilde F}_0}{\partial u}\right|_{u_\phi}\right)+\frac{i k^2 M_0 \chi}{2 n_0 q}\, ,
\label{disp4}
\end{equation}
where, neglecting terms of order $(\partial\tilde F / \partial u)|^2_{u_\phi} $, the real part of the frequency is given by

\begin{equation}
  \label{disp5}
  \omega_r=\omega_p\left(1 - \frac{M_0\chi\pi\omega_p}{2 q n_0} \left.\frac{\partial{\tilde F}_0}{\partial u}\right|_{u_\phi}\right)\,,
\end{equation}
and $M_0=n_0\mu_e\eta(\alpha)$ is value of the spin
magnetization of the system. This is because the distribution $f_0$ of
Eq.~\eqref{distribu0}, gives the zeroth order magnetization of the
system ${\bf M}_0=(2\mu_e n_0/\hbar)\int{\bf s} f_0d{\bf v}{d\bf s}=M_0\hat z$
\cite{marklund,brodin}. Thus, using Eqs.~\eqref{disp4} and
\eqref{disp5} the imaginary part of the frequency is
\begin{equation}
  \label{disp6}
  \omega_i=\frac{\pi \omega_p^3}{2 k^2}\left.\frac{\partial{\tilde F}_0}{\partial u}\right|_{u_\phi}+\frac{k^2 M_0 \chi}{2 n_0 q}\, ,
\end{equation}

From \eqref{disp6} we note that there is a correction in the
imaginary part of the frequency which appears due to the magnetization of
the plasma due to spin of their constituents. This correction depends
on the electromagnetic part of the waves, which is coupled with spin
vector of each particle. As we discuss in our previous
work~\cite{faz}, the interaction of spin with the perturbed magnetic
field in a magnetized plasma is the responsible of this contribution
to the damping of an electrostatic mode. Moreover, in the present
analysis we show that the energy trasfer between waves and particles
depends on the shape of the equilibrium function (classical Landau
damping) and also on the spin magnetization of the plasma. 

The spin damping correction of Eq.~\eqref{disp6} depends on
the ratio of the magnetization and the charge of each
particle, i.e, $M_0/n_0q = \hbar g\eta(\alpha)/4m$. It is expected that spin
effects will be important at low temperatures, high densities and huge
magnetic fields, and due to the dependences of $\eta$ on the $\alpha$ parameter we
can see that, in fact, the spin correction to the Landau damping is
higher for high values of density and background magnetic field, and
for low temperatures. However, this spin damping
is proportional to $\hbar$ and the main effect is the classical Landau
damping. 

Besides, we can see that the correction to the damping is proportional to $\chi$, which
depends on the transversal part of the electromagnetic wave. 
If $\chi>0$ in the case of an electron plasma, $\eta(\alpha)<0$ and the correction
is a damping. Also, for the same electron plasma, if $\chi<0$ the
correction is an instability. 
When the wave has no electromagnetic transverse component $|{\bf
  E_{\perp}}|=0$, then $\chi=0$ and there are no spin correction to the
damping. The exact value of this coefficient should be obtained solving the dynamical
Maxwell equations with the current density  ${\bf j}=qn_0\int {\bf v} f_1
d{\bf v}d{\bf s}$. The complete implications of the value of $\chi$ is being studied.

On the other hand, from Eq.~\eqref{disp5} we note that there is a
correction in the frequency of plasma oscillations. This corrections
depends on the magnetization of the plasma and also on the shape of
the distribution function. As spin corrections are proportional to $\hbar$
and the derivative of the distribution function is small, as in the
case of the imaginary part, the correction is small and the frequency
of electron waves is near $\omega_p$ as expected.

In conclusion, we have shown that the incorporation of spin to kinetic
theory in a magnetized plasma produces corrections to the classical
Landau damping for electrostatic waves that depends on the
magnetization of the system, and is due to the coupling between the
spin vector and the electromagnetic part of the wave. In other words,
in addition to the Landau mechanism to transfer energy from waves to
particles, the inclusion of spin allows the energy transfer through
the quantum interaction between spin and magnetic fields. However, the
corrections are of $\hbar$ order and, in the case of electron plasma and mawellian
distribution functions, the electrostatic wave will show an evolution
similiar to its classical dynamics when the spin is not included.
In addition we shown that spin contribution also introduces a
correction to the frequency of plasma oscillations which is also
proportional to $\hbar$ and it is due to the magnetization of the plasma.
 
All of these results show the importance of the kinetic theory of
plasmas. The same formalism can be used to introduce other quantum
contributions to classical plasma physics and derive new corrections
and effects for strongly coupled plasmas, as well as plasmas in
presence of large magnetic fields when quantum effects are relevant.


Pablo S. Moya is grateful to CONICyT D-21070397 Doctoral Fellowship.


\end{document}